\input harvmac
\overfullrule=0pt
\parindent 25pt
\tolerance=10000

\input epsf

\newcount\figno
\figno=0
\def\fig#1#2#3{
\par\begingroup\parindent=0pt\leftskip=1cm\rightskip=1cm\parindent=0pt
\baselineskip=11pt
\global\advance\figno by 1
\midinsert
\epsfxsize=#3
\centerline{\epsfbox{#2}}
\vskip 12pt
{\bf Fig.\ \the\figno: } #1\par
\endinsert\endgroup\par
}
\def\figlabel#1{\xdef#1{\the\figno}}
\def\encadremath#1{\vbox{\hrule\hbox{\vrule\kern8pt\vbox{\kern8pt
\hbox{$\displaystyle #1$}\kern8pt}
\kern8pt\vrule}\hrule}}

 \def\ep{{\epsilon}}

 \def\frac#1#2{{#1\over #2}}

 \def\s{\sqrt}

 \def\al{\alpha'}
 \def\de{\partial}
 
 \def\we{\wedge}
 
 \def\f {\frac}
 \def\ti{\tilde}
 \def\ap{\alpha}

 \def\ddd{\cdot\cdot\cdot}
 
 \def\la{\langle}
 \def\lb{\rangle}
 \def\ep{\epsilon}

 \def\vp{\varphi}

\lref\tds{
D.~J.~Gross and N.~Miljkovic,
``A Nonperturbative Solution Of D = 1 String Theory,''
Phys.\ Lett.\ B {\bf 238}, 217 (1990);

E.~Brezin, V.~A.~Kazakov and A.~B.~Zamolodchikov,
``Scaling Violation In A Field Theory Of Closed Strings
In One Physical Dimension,''
Nucl.\ Phys.\ B {\bf 338}, 673 (1990);

P.~Ginsparg and J.~Zinn-Justin,
``2-D Gravity + 1-D Matter,''
Phys.\ Lett.\ B {\bf 240}, 333 (1990).
}

\lref\GRa{ D.~Gaiotto and L.~Rastelli, ``A paradigm of open/closed
duality: Liouville D-branes and the Kontsevich model,''
arXiv:hep-th/0312196.
}

\lref\DK{P.~Di Francesco and D.~Kutasov,
``World sheet and space-time physics in two-dimensional (Super)string
theory,''
Nucl.\ Phys.\ B {\bf 375}, 119 (1992)
[arXiv:hep-th/9109005].
}

\lref\GTT{ S.~Gukov, T.~Takayanagi and N.~Toumbas, ``Flux
backgrounds in 2D string theory,'' arXiv:hep-th/0312208.
}

\lref\MV{ J.~McGreevy and H.~Verlinde, ``Strings from tachyons:
The c = 1 matrix reloaded,'' [arXiv:hep-th/0304224].
}

\lref\KMS{ I.~R.~Klebanov, J.~Maldacena and N.~Seiberg, ``D-brane
decay in two-dimensional string theory,'' [arXiv:hep-th/0305159].}

\lref\TT{T.~Takayanagi and N.~Toumbas, ``A matrix model dual of
type 0B string theory in two dimensions,'' JHEP {\bf 0307}, 064
(2003) [arXiv:hep-th/0307083].
}

\lref\KlR{I.~R.~Klebanov, ``String theory in two-dimensions,''
[arXiv:hep-th/9108019].}

\lref\six{M.~R.~Douglas, I.~R.~Klebanov, D.~Kutasov, J.~Maldacena,
E.~Martinec and N.~Seiberg, ``A new hat for the c = 1 matrix
model,'' [arXiv:hep-th/0307195].
}

\lref\JY{A.~Jevicki and T.~Yoneya, ``A Deformed matrix model and
the black hole background in two-dimensional string theory,''
Nucl.\ Phys.\ B {\bf 411}, 64 (1994) [arXiv:hep-th/9305109].
}

\lref\Sen{ A.~Sen, ``Open-closed duality: Lessons from matrix
model,'' [arXiv:hep-th/0308068].
}
\lref\Ka{A.~Kapustin, ``Noncritical superstrings in a
Ramond-Ramond background,'' [arXiv:hep-th/0308119].
}

\lref\ADKMV{ M.~Aganagic, R.~Dijkgraaf, A.~Klemm, M.~Marino and
C.~Vafa, ``Topological strings and integrable hierarchies,''
arXiv:hep-th/0312085.
}

\lref\ES{ T.~Eguchi and Y.~Sugawara, ``Modular bootstrap for
boundary N = 2 Liouville theory,'' JHEP {\bf 0401}, 025 (2004)
[arXiv:hep-th/0311141].
}

\lref\DKR{ K.~Demeterfi, I.~R.~Klebanov and J.~P.~Rodrigues, ``The
Exact S matrix of the deformed c = 1 matrix model,'' Phys.\ Rev.\
Lett.\  {\bf 71}, 3409 (1993) [arXiv:hep-th/9308036].
}

 \lref\MP{ E.~Marinari and G.~Parisi, ``The Supersymmetric
One-Dimensional String,'' Phys.\ Lett.\ B {\bf 240}, 375 (1990).
}

\lref\MMVR{ J.~McGreevy, S.~Murthy and H.~Verlinde,
``Two-dimensional superstrings and the supersymmetric mat rix
model,'' arXiv:hep-th/0308105.
}

\lref\FH{T.~Fukuda and K.~Hosomichi, ``Three-point functions in
sine-Liouville theory,'' JHEP {\bf 0109}, 003 (2001)
[arXiv:hep-th/0105217].
}

\lref\SA{Y.~Satoh, ``Three-point functions and operator product
expansion in the SL(2)  conformal field theory,'' Nucl.\ Phys.\ B
{\bf 629}, 188 (2002) [arXiv:hep-th/0109059].
}

\lref\seibergR{
N.~Seiberg,
``Notes On Quantum Liouville Theory And Quantum Gravity,''
Prog.\ Theor.\ Phys.\ Suppl.\  {\bf 102}, 319 (1990).
}

\lref\GK{D.~J.~Gross and I.~R.~Klebanov,
``Fermionic String Field Theory Of C = 1 Two-Dimensional Quantum Gravity,''
Nucl.\ Phys.\ B {\bf 352}, 671 (1991).
}

\lref\DR{
K.~Demeterfi and J.~P.~Rodrigues,
``States and quantum effects in the collective field theory of a deformed
matrix model,''
Nucl.\ Phys.\ B {\bf 415}, 3 (1994)
[arXiv:hep-th/9306141].
}

\lref\GRO{
P.~Bouwknegt, J.~G.~McCarthy and K.~Pilch,
``Ground ring for the 2-D NSR string,''
Nucl.\ Phys.\ B {\bf 377}, 541 (1992)
[arXiv:hep-th/9112036].
}

\lref\BMP{
P.~Bouwknegt, J.~G.~McCarthy and K.~Pilch,
``BRST analysis of physical states for 2-D (super)gravity coupled to
(super)conformal matter,''
arXiv:hep-th/9110031.
}

\lref\IO{
K.~Itoh and N.~Ohta,
``BRST cohomology and physical states in 2-D
supergravity coupled to $c \leq 1$
matter,''
Nucl.\ Phys.\ B {\bf 377}, 113 (1992)
[arXiv:hep-th/9110013];
``Spectrum of two-dimensional (super)gravity,''
Prog.\ Theor.\ Phys.\ Suppl.\  {\bf 110}, 97 (1992)
[arXiv:hep-th/9201034].
}

\lref\INOT{
H.~Ita, H.~Nieder, Y.~Oz and T.~Sakai,
``Topological B-model, matrix models, c-hat = 1 strings and quiver gauge
theories,''
JHEP {\bf 0405}, 058 (2004)
[arXiv:hep-th/0403256].
}

\lref\KMSG{
D.~Kutasov, E.~J.~Martinec and N.~Seiberg,
``Ground rings and their modules in 2-D gravity with $c \leq 1$ matter,''
Phys.\ Lett.\ B {\bf 276}, 437 (1992)
[arXiv:hep-th/9111048].
}

\lref\WiG{
E.~Witten,
``Ground ring of two-dimensional string theory,''
Nucl.\ Phys.\ B {\bf 373}, 187 (1992)
[arXiv:hep-th/9108004].
}

\lref\five{
O.~DeWolfe, R.~Roiban, M.~Spradlin, A.~Volovich and J.~Walcher,
``On the S-matrix of type 0 string theory,''
JHEP {\bf 0311}, 012 (2003)
[arXiv:hep-th/0309148].
}

\lref\Po{
A.~M.~Polyakov,
``Quantum Geometry Of Fermionic Strings,''
Phys.\ Lett.\ B {\bf 103}, 211 (1981).
}

\lref\DHK{
J.~Distler, Z.~Hlousek and H.~Kawai,
``Superliouville Theory As A Two-Dimensional, Superconformal Supergravity
Theory,''
Int.\ J.\ Mod.\ Phys.\ A {\bf 5}, 391 (1990).
}

\lref\Da{
U.~H.~Danielsson,
``A matrix model black hole: Act II,''
JHEP {\bf 0402}, 067 (2004)
[arXiv:hep-th/0312203].
}

\lref\BGV{
N.~Berkovits, S.~Gukov and B.~C.~Vallilo,
``Superstrings in 2D backgrounds with R-R flux and new extremal black
holes,''
Nucl.\ Phys.\ B {\bf 614}, 195 (2001)
[arXiv:hep-th/0107140].
}

\lref\THM{ J.~Davis, L.~A.~Pando Zayas and D.~Vaman, ``On black
hole thermodynamics of 2-D type 0A,'' JHEP {\bf 0403}, 007 (2004)
[arXiv:hep-th/0402152];
U.~H.~Danielsson, J.~P.~Gregory, M.~E.~Olsson, P.~Rajan and
M.~Vonk, ``Type 0A 2D black hole thermodynamics and the deformed
matrix model,'' JHEP {\bf 0404}, 065 (2004)
[arXiv:hep-th/0402192].
}

\lref\KRE{
D.~Kutasov,
``Some properties of (non)critical strings,''
arXiv:hep-th/9110041.
}

\lref\St{
A.~Strominger,
``A matrix model for AdS(2),''
JHEP {\bf 0403}, 066 (2004)
[arXiv:hep-th/0312194].
}

\lref\HV{
H.~Verlinde,
``Superstrings on AdS(2) and superconformal matrix quantum mechanics,''
arXiv:hep-th/0403024.
}

\lref\NTT{
T.~Nakatsu, K.~Takasaki and S.~Tsujimaru,
``Quantum and classical aspects of deformed c = 1 strings,''
Nucl.\ Phys.\ B {\bf 443}, 155 (1995)
[arXiv:hep-th/9501038].
}

\lref\MPR{G.~W.~Moore, M.~R.~Plesser and S.~Ramgoolam,
``Exact S matrix for 2-D string theory,''
Nucl.\ Phys.\ B {\bf 377}, 143 (1992)
[arXiv:hep-th/9111035].
}

\lref\AH{
K.~i.~Aoki and E.~D'Hoker,
``Correlation functions of minimal models coupled to two-dimensional quantum
supergravity,''
Mod.\ Phys.\ Lett.\ A {\bf 7}, 333 (1992)
[arXiv:hep-th/9109025].
}

\lref\DKB{
P.~Di Francesco and D.~Kutasov,
``Correlation functions in 2-D string theory,''
Phys.\ Lett.\ B {\bf 261}, 385 (1991).
}

\lref\Ta{
T.~Takayanagi,
``Notes on D-branes in 2D type 0 string theory,''
JHEP {\bf 0405}, 063 (2004)
[arXiv:hep-th/0402196].
}

\lref\Poi{ S.~H.~Shenker, ``The Strength Of Nonperturbative
Effects In String Theory,'' RU-90-47; J.~Polchinski,
``Combinatorics Of Boundaries In String Theory,'' Phys.\ Rev.\ D
{\bf 50}, 6041 (1994) [arXiv:hep-th/9407031].
}

\lref\OM{
E.~Martinec and K.~Okuyama,
``Scattered Results in 2D String Theory,''
arXiv:hep-th/0407136.
}

\lref\GM{
P.~H.~Ginsparg and G.~W.~Moore,
``Lectures on 2-D gravity and 2-D string theory,''
arXiv:hep-th/9304011.
}

\lref\PoR{
J.~Polchinski,
``What is string theory?,''
arXiv:hep-th/9411028.
}

\lref\Nak{
Y.~Nakayama,
``Liouville field theory: A decade after the revolution,''
arXiv:hep-th/0402009.
}

\lref\GW{
D.~J.~Gross and J.~Walcher,
``Non-perturbative RR potentials in the c(hat) = 1 matrix model,''
JHEP {\bf 0406}, 043 (2004)
[arXiv:hep-th/0312021].
}

\baselineskip 18pt plus 2pt minus 2pt

\Title{\vbox{\baselineskip12pt
\hbox{hep-th/0408086}\hbox{HUTP-04/A0047}
  }}
{\vbox{\centerline{Comments on 2D Type IIA String and Matrix
Model}}} \centerline{Tadashi Takayanagi\foot{e-mail:
takayana@bose.harvard.edu}}

\medskip\centerline{ Jefferson Physical Laboratory}
\centerline{Harvard University}
\centerline{Cambridge, MA 02138, USA}

\vskip .1in \centerline{\bf Abstract} We consider a type
IIA-like
string theory with RR-flux in two dimension and propose its matrix
model dual. This string theory describes a Majorana fermion in the
 two dimensional spacetime. We also discuss its
scattering amplitudes both in the world-sheet theory
and in the matrix model.

\noblackbox

\Date{}

\writetoc

\newsec{Introduction}

The two dimensional string theory \tds\ is a very useful
laboratory of quantum gravity, where we can solve the theory
exactly by employing the dual $c=1$ matrix model\foot{For reviews
see e.g. \KlR \GM \PoR \Nak.}. In spite of its low dimensionality
this string theory is dynamical by exciting a massless scalar
field. Thus we can analyze the dynamical processes in the two
dimensional quantum gravity non-perturbatively. Even though the
two dimensional bosonic string turned out to be non-perturbatively
unstable \Poi, the two dimensional type 0 string is
non-perturbatively well-defined \TT \six . This can be seen from
its matrix model dual \TT \six , obtained
  in the light of the recent interpretation of
fermions in the matrix model as unstable D-branes \MV \KMS.

Usually in superstring we have two kinds of theories i.e. type II
and type 0 , and we are more interested in type II than type 0
since the latter includes a closed string tachyon and no fermions
in ten dimension. Thus it will be basic and important to ask
whether we can construct a non-perturbatively stable type II
string in two dimension, though even the type 0 theory is
completely stable in two dimension. One way to define a type II
string is to take a ${\bf Z}_2$ orbifold of type 0 string by the
action $(-1)^{F_L}$ of the world-sheet fermion parity, though this
theory has no spacetime supersymmetry. In type II string we cannot
put a cosmological constant term to make the theory weakly coupled
because it is not allowed by the GSO projection. Instead we can
consider a similar Liouville-like term of RR vertex operators. In
the paper \GTT\ a matrix dual of type IIB-like string theory was
proposed following this idea (see \MMVR \HV\ for other proposals
by using the supersymmetric matrix model \MP ). The resulting
structure is the same as that of the $c=1$ matrix model. However,
this construction seems to be non-perturbatively unstable as noted
in \GTT\ and we will probably need a refinement about its
non-perturbative corrections.

Motivated by this, we would like to consider a type IIA-like
string (below we will just call this a type IIA string) from the
viewpoints of both the world-sheet and matrix model. We consider
type 0A with RR-flux (see e.g. \Ka  \GW \St \Da \GTT\ for recent
discussions) and take a ${\bf Z}_2$ quotient to define the type
IIA string. As we will see later, the dynamical field in the IIA
string is a Majorana fermion coupled to the two dimensional
linear-dilaton gravity. Obviously this model is non-perturbatively
well-defined as in the original 0A theory.

We organize this paper as follows. In section 2 we give a
world-sheet description of two dimensional type II string. In
section 3 we propose a matrix model dual of type IIA string and
discuss its properties. In section 4 we compute some of tree level
scattering amplitudes in the IIA string and try to compare them
with the results in the dual matrix model. In section 5 we
summarize the conclusions.

\newsec{World-sheet Theory of 2D Type IIA String}

\subsec{Definition of 2D Type II String}

The world-sheet fields\foot{In this paper we set $\alpha '=2$.} in
two dimensional superstring consist of the $\hat{c}=1$ matter and
the super-Liouville fields \Po \DHK \DK. The fields in the
$\hat{c}=1$ sector are a free boson $X_0$ and its superpartner
$\psi_0$. They describe the time coordinate of the two dimensional
spacetime. Those in the super-Liouville sector are the Liouville
field $\phi$ and its superpartner $\psi_{1}$. The Liouville field
has the background charge $Q=2$ (central charge $c=1+3Q^2=13$) and
describes the space coordinate with a linear dilaton
$g_s=e^{\phi}$. Since the string theory becomes strongly coupled
when $\phi$ becomes large, usually we put a (super)Liouville term
\eqn\slt{ \mu \int dz^2 d\theta^2 e^{\Phi},} in order to regulate
the strongly coupled region. Then we can define physical vertex
operators in NS and R sectors by
\eqn\pver{\eqalign{V_{NS}&=e^{-\vp}e^{\phi+iP(\phi\pm X^0)}, \cr
V_{R(\ep)}&=e^{-\f{1}{2}\vp}e^{\ep\f{i}{2}H}e^{\phi+ iP( \phi+\ep
X^0)},}} where $\ep=\pm$ represents the chirality of R-sector
fermion in the spacetime and comes from the two choices of ground
states in R-sector. The field $H$ is the bosonization of the two
real fermions $\psi_0$ and $\psi_1$. Note that the chirality
determines the traveling direction of R-sector fields due to the
`Dirac equation' or the super-Virasoro constraint
$G_{0}|phys\lb=0$. We call a field with $\ep=+$ (or $\ep=-$) a
left-moving (or right-moving) one in the two dimensional
spacetime.

To define consistent string models we need GSO projections. The
non-chiral GSO projections lead to the type 0A and 0B theory \TT
\six\ defined by the following chiral- and antichiral- sectors of
closed string, \eqn\sect{\eqalign{0A:& (NS,NS),(R(+),R(-)),
(R(-),R(+)),\cr
          0B:&  (NS,NS),(R(+),R(+)), (R(-),R(-)).}}
The NSNS sector in each theory represents a massless scalar field,
which was originally a tachyon field in the familiar ten
dimensional type 0 string theory. The RR sector in the 0B theory
corresponds to a scalar field (axion). In the 0A theory a RR
vertex operator cannot have a non-zero momentum. This is because
it corresponds to a one-form RR gauge potential whose
field-strength is two form in the two dimensional spacetime. Its
equation of motion requires that the RR-flux should be constant.
The matrix model dual to these theories were given\foot{A further
check of this matrix model proposal of the 2d type 0 string was
given in \Ta\ from the viewpoint of holography (or open/closed
duality) by analyzing loop operators. From this analysis we can
find the world-sheet supersymmetry implicitly by identifying what
are the NSNS and RR sector in the matrix model side, which is the
most important difference from the 2d bosonic string. For other
related discussions on the interpretations of $c=1$ or $c<1$
matrix models via open/closed duality, refer to \MV \Sen \ADKMV
\GRa.} in \TT \six.

In this paper we would like to consider other kinds of two
dimensional string models obtained from a chiral GSO-projection.
We call them type IIA and IIB, since they can also be obtained
from the ${\bf Z}_2$ orbifold by the action $(-1)^{F_L}$ of the 0A
and 0B theory. These type II models
 are defined by the sectors
\eqn\sectii{\eqalign{IIA:& (NS,R(-)),(R(+),NS), (R(-),R(+)),\cr
          IIB:&  (NS,R(-)),(R(-),NS), (R(+),R(+)).}}
The chiralities of R-sectors are determined such that the OPEs
between vertex operators are local with each other, which is a
usual procedure to find a correct GSO projection. Earlier
discussions of the related typeII-like models can be found in  \DK
\KRE \MMVR. Indeed the model \sectii\ we are disucssing here is
equivalent to the zero radius limit $R\to 0$ of the two
dimensional superstring considered in \KRE. In the paper \KRE, the
superstring model is defined to be the ${\bf Z}_2$ orbifold of
compactified type 0 string by the ${\bf Z}_2$ action
$(-1)^{F_L}\cdot\sigma_{1/2}$, where $\sigma_{1/2}$ denotes the
half-shift $X\to X+\pi R$. Thus in the limit $R\to 0$ the model
becomes\foot{ Notice that the orbifold action (3.23) in \KRE\ is
the same as $(-1)^{F_L}\sigma_{1/2}$. Indeed we can check that the
explit spectrum (3.24),(3.25) and (3.26) obtained in \KRE\ leads
to only fields in NSR and RR sectors in the limit $R\to 0$ because
the mass becomes infinite in the NSNS sector. On the other hand,
when $R\to \infty$, the NSR sector fields become infinitely
massive as is expected in type 0 string.} type II string and is
the same as our non-compact model \sectii\ after T-duality.

The physical fields in the IIB background are given by a
left-moving RR scalar field and also a right-moving Dirac fermion
in the NSR and RNS sector. Though we call this theory type II,
there is no spacetime supersymmetry actually as long as the
time-direction is not compactified. Applying a bosonization of the
fermion in two dimension, we get a single massless scalar field
after combined with the RR field. This is the same field content
as the familiar two dimensional bosonic string. Indeed as argued
in \GTT, the matrix model dual of IIB can be obtained from the
corresponding
 ${\bf Z}_2$ projection of type 0B model with a non-zero RR-flux in the
 matrix model side\foot{See also \MMVR\ for another proposal
 for a N=2 Liouville NSNS
background using a supersymmetric matrix model \MP\ from the
viewpoint of multiples D-branes. The D-branes in the N=2 Liouville
theory were classified in \ES.} and this has the same structure as
the $c=1$ matrix model. In this definition, a constant RR-flux
plays a role of the `Liouville term' which regulates the strongly
coupled region\foot{More precisely, we should say that the
Liouville like term consists of RR-flux and also bosonized field
of NSR fermion as discussed in \GTT.} instead of the conventional
cosmological constant term \slt. Notice that the term \slt\
corresponds to a closed string tachyon condensation and the
tachyon field
 is projected out in type II models. However,
this definition of IIB string is at the perturbative level since
the RR-flux background of the original type 0B theory is not
non-perturbatively well defined. It is not clear how to define the
matrix model dual to type IIB non-perturbatively, though we
believe that should be possible.

Motivated by this we would like to turn to the IIA model\foot{ A
classical Green-Schwarz string of type IIA string $AdS_2$ was
proposed recently in \HV\ by using a supercoset (see also
\BGV).} because the
RR-flux background in the 0A model is non-perturbatively
well-defined. In the IIA model, the spacetime fields in the
$(R(+),NS)$ and $(NS,R(-))$ sectors are given by a Majorana
left-moving and right-moving fermion $\psi(x^+)$ and
$\ti{\psi}(x^-)$. The RR field in $(R(-),R(+))$ is RR 1-form
potential $C$ (or 2-form RR field-strength $F$). Note that the
physical state constraint $G_{0}|phys\lb=0$ kills all propagating
modes and only the zero-mode is allowed as in the 0A case. Again
we have the background constant RR-flux $F_{t\phi}=q$, which is
represented by the vertex operator
 \eqn\rrl{S_{RRflux}=q\int_{\Sigma}
dzd\bar{z}\ V_{RR}=
 q \int_{\Sigma}
dzd\bar{z}\  e^{-\f{1}{2}\vp(z)-\f{1}{2}\vp(\bar{z})}
e^{\f{i}{2}H(z)-\f{i}{2}H(\bar{z})} e^{\phi(z,\bar{z})}.}
This includes an exponential $e^{\phi}$ factor and thus
regulates the strongly coupled region at $\phi\sim -\log q$.
The large $q$ means that the theory is weakly coupled. The effective
description by the extremal black-hole solution \BGV\
will also be applied to this IIA
model with RR-flux as in the 0A model discussed in \Da \GTT\ (see also
\THM\ \OM\ for further discussions).

\subsec{Minisuperspace Approximation}

As we have seen, the background RR-flux $F_{t\phi}=q$ makes the
IIA theory weakly coupled. It will lead to an effective potential
wall in the same way as \slt. The reflection of a propagating
field due to such a wall can be usually well described by the
minisuperspace approximation as has been done in bosonic string
\seibergR\ and type 0 string \six. Since the RR-vertex operator is
also proportional to $e^{\phi}$ (Liouville dressing) in our case,
we may expect the minisuperspace approximation is given by the
 action like
\eqn\maction{S=\int dt d\phi\left[i\psi\de_{+}\psi
+i\ti{\psi}\de_{-}\ti{\psi}+2iq e^{\phi}\psi\ti{\psi}\right],}
where $x^{\pm}=\f{1}{2}(t\pm \phi)$. The Majorana fermions $\psi$
and $\ti{\psi}$ correspond to the left-moving and right-moving
part in the asymptotic region $\phi\to -\infty$, respectively.
This can be understood if we assume the (Lorentz invariant)
coupling $\bar{\Psi} F_{\mu\nu}\Gamma^{\mu\nu}\Gamma^{01}\Psi$
between the fermions and RR-fields. We can also find the similar
behavior from the matrix model side discussed later. We can solve
the wavefunction of \maction\ exactly as follows. The equation of
motion is given by \eqn\emo{\de_{+}\psi+2q\ti{\psi}e^{\phi}=0, \ \
\ \ \de_{-}\ti{\psi}-2q\psi e^{\phi}=0.} The Majorana fermions
$\psi$ and $\ti{\psi}$ with energy $\omega$ satisfy the
differential equations ($l=e^{\phi}$) \eqn\diff{\eqalign{ &
(l^2\de_{l}^2+\omega^2+ i\omega-4q^2 l^2)\psi=0, \cr &
(l^2\de_{l}^2+\omega^2- i\omega-4q^2 l^2)\ti{\psi}=0.}} Also we
require that in the limit $\phi\to +\infty$ they should decay
exponentially due to the mass term or Liouville like potential.
Thus we find the solutions up to a normalization
\eqn\soldf{\eqalign{ & \psi(t,\phi)= e^{-i\omega
t}e^{\phi/2}K_{i\omega-1/2}(2q e^{\phi}), \cr
&\ti{\psi}(t,\phi)=e^{-i\omega t}e^{\phi/2}K_{i\omega+1/2}(2q
e^{\phi}).}} Indeed they behave in the strongly coupled region
$\phi\to +\infty$ as \eqn\stron{\ti{\psi}(t,\phi)\sim
{\psi}(t,\phi)\sim \f{\pi}{4q}e^{-i\omega t}e^{-q e^\phi}.} On the
other hand in the weakly coupled region $\phi\to -\infty$,
\eqn\soldfg{\eqalign{ & \psi(t,\phi)=\f{1}{2}q^{i\omega-1/2}
\Gamma\left(\f{1}{2}-i\omega\right)\ e^{-i\omega (t-\phi)}, \cr
 & \ti{\psi}(t,\phi)=\f{1}{2} q^{-i\omega-1/2}
\Gamma\left(\f{1}{2}+i\omega\right)\ e^{-i\omega (t+\phi)}.}}
{}From this we find the reflection amplitude of NSR fermion in the
minisuperspace approximation
\eqn\legfm{S(\omega)^{ms}_{NSR}=q^{-2i\omega}
\f{\Gamma\left(\f{1}{2}+i\omega\right)}
{\Gamma\left(\f{1}{2}-i\omega\right)}.} Interestingly, this result
is the same as the minisuperspace approximation of the RR scalar
field in the type 0B matrix model computed in \six.

\subsec{Discrete States}

In addition to the continuous state \pver, there are also another
kind of physical states at discrete imaginary momenta. This is a
special property of two dimensional string theories. They are
called discrete states and are computed in \IO \BMP\ in the
$\hat{c}=1$ case.
 In a chiral sector they are given by
(for positive $r$ and $s$) \eqn\discr{ |W_{(r,s)}\lb= \left(\int
\psi e^{-iX}\right)^s \
e^{\f{i}{2}(r+s)X}e^{\f{1}{2}(r+s+2)\phi}|0\lb,} in the Euclidean
theory ($X=iX_0$), where $r$ and $s$ are integers which satisfy
$rs>0$. The states for negative $r$ and $s$ can also be obtained
in a similar way by a dual operation. The state $|0\lb$ represents
the ground state i.e. the -1 picture vacuum  for the NS sector and
-1/2 picture vacuum with the negative chirality for the R sector.
Discrete states with $r-s\in 2{\bf Z}$ (or $r-s\in 2{\bf Z}+1$)
belong to the NS-sector (or R-sector). Notice that this fact is
consistent with the position of poles of the NSNS and RR leg
factor $e^{\delta_{NSNS}(p)}\propto\f{\Gamma(iP)}{\Gamma(-iP)}$
and $e^{\delta_{RR}(P)}\propto\f{\Gamma(1/2+iP)}{\Gamma(1/2-iP)}$
\TT \six, respectively. When we consider the chiral and antichiral
sector to define a closed string theory, the momentum of $\phi$
should be the same in both sectors. Thus the discrete states
appear only in NSNS and RR sectors. Furthermore, since here we
assume that the time coordinate is also non-compact, the momentum
of $X$ should also be the same in the chiral and antichiral
sector.

The 0B theory includes all of the NSNS and RR discrete states in
the left-right symmetric way. For example, a massive graviton
appears in the NSNS discrete sates. In the 0A theory discrete
states only exist in the NSNS sector. There are no RR-sector ones
because of the left-right asymmetric GSO projection. In a similar
way we can also define the discrete states in the type IIA and IIB
model. Both can be obtained after the ${\bf Z}_2$ twist of the 0A
and 0B by the operator $(-1)^{F_L}$. In IIB theory they are given
by $(r,s)\in (2{\bf Z}+1,2{\bf Z}+1)$ for the NSNS-sector and
$(r,s)\in  (2{\bf Z},2{\bf Z}+1)$ for the RR-sector. In the IIA
model we get $(r,s)\in  (2{\bf Z}+1,2{\bf Z}+1)$ for the
NSNS-sector and none for the RR-sector. Note that there are no
twisted sectors (or equally the NSR and RNS sector) as there are
no discrete states in the NSR or RNS sector.

Finally we would like to briefly examine the ground ring structure
of these string models. We define the generators of the ground
ring by $x$ and $y$. Each of them is a BRST invariant operator
with ghost number zero including both chiral and antichiral part
symmetrically. For the precise definition of these operators see
\GRO \six. In the type 0B the ground ring is generated by $x$ and
$y$, while in the 0A it is generated by $x^2,y^2$ and $xy$. The
discrete state $W_{(r,s)}$ corresponds to the ground ring element
$x^{-r-1}y^{-s-1}$ \GRO. The ground ring structure of type 0
string was discussed from geometrical viewpoints in \INOT. We
would also like to apply this argument to the type II models. The
ground ring of IIA is generated by $x^2$ and $y^2$, while that of
IIB by $x$ and $y^2$. This result can be easily understood as the
${\bf Z}_2$ projection $y\to -y$, which is equivalent to
$(-1)^{F_L}$. Generally the ground ring structure represents the
$W_{\infty}$ symmetry of matrix model \WiG. As we will see later,
the above results on type II string can be correctly reproduced
from the proposed matrix model.

\newsec{A Proposal of Matrix Model Dual}

\subsec{IIA Matrix Model}

Now we would like to construct a matrix model dual of the type IIA
string in two dimension, whose properties on the world-sheet have
been discussed in the previous section. We argue that the IIA
matrix model can be obtained by the ${\bf Z}_2$ projection
$(-1)^{F_R}=-(-1)^{F_L}=1$ of the type 0A model as in the
world-sheet theory\foot{A matrix model dual of IIA string in
$AdS_2$ was proposed in \HV. As is clear from section 2.3, our
model does not have any super-$W_{\infty}$ symmetry, while the
$AdS_2$ model seems to have it.}. The similar construction has
already proposed for IIB model in \GTT. The IIB model is defined
by the ${\bf Z}_2$ projection $(-1)^{F_L}=1$ of the 0B matrix
model. This ${\bf Z}_2$ action can be identified with
 the transformation of a fermion (or hole) into a hole (or fermion)
and the operation $(x,p)\to (p,x)$ at the same time \six. The
cosmological constant changes its sign under this action. We will
apply the similar method to define the IIA model.

The 0A matrix model \six \TT\ is equivalent to the Hermitian
matrix model with the following deformed Hamiltonian \JY\ as shown
in \Ka \eqn\pota{2H=p^2-x^2+\f{M}{x^2},\ \ \ M\equiv
q^2-\f{1}{4}.} In addition to the Hamiltonian, we have the
following conserved charges \DR\ almost in the same way as the
$W_{\infty}$ charge in the $c=1$ matrix model \eqn\winf{\eqalign{
&W_{+}=e^{-2t}\left( (p+x)^2+\f{M}{x^2}\right)=2\s{M+\mu^2}, \cr
&W_{-}=e^{2t}\left((p-x)^2+\f{M}{x^2}\right) =2\s{M+\mu^2},}} for
the classical trajectory $x^2(t)=\mu+\s{M+\mu^2}\cosh(2t)$. Any
general conserved charges can be written by the product of
$W_{+},W_{-}$ and $H$ summed over each fermions \DR. In the 0A
theory we can also define the ${\bf Z}_2$ action by the
combination of the hole-particle exchange and the action
\eqn\acty{x'= \s{p^2+\f{M}{x^2}},\ \ \
p'=\f{px}{\s{p^2+\f{M}{x^2}}}.} Under this transformation, we can
show $dx'\we dp'=-dx\we dp$ and thus this is a canonical
transformation. This extra minus sign, which also appears in the
0B case (just setting $M=0$), is due to the fact that the hole has
a minus momentum compared with a particle. Under this ${\bf Z}_2$
action, the Hamiltonian changes its sign $H\to -H$ (or equally
$\mu\to -\mu$), while the other charges remain unchanged
$W_{\pm}\to W_{\pm}$. These facts are consistent with our previous
analysis on the ground ring structure. As in the usual $c=1$
matrix model, the three ground ring generators $x^2,y^2$ and $xy$
of 0A model are naturally identified with $W_{+},W_{-}$ and $H$,
respectively. Only the operator $xy$ changes its sign under the
${\bf Z}_2$ action by $(-1)^{F_L}$ and it is projected out in IIA
model. This exactly agrees with its action in our matrix model
side. The same argument can also be applied to the matrix model
dual of IIB theory \GTT. We can also see this in a
non-perturbative way by applying the exact wave function \DKR\ to
the quantum mechanics for \pota. Indeed the density of
state\foot{This can be computed from the phase shift of the wave
function.} for the fermions is ${\bf Z}_2$ symmetric
$\rho(\ep,q)=\rho(-\ep,q)$ \GTT. Then the non-perturbative ${\bf
Z}_2$ action can be written as \eqn\actionn{a_{\ep} \to
a^{\dagger}_{-\ep},} where we denote the creation and annihilation
operator of fermion with the energy $\ep$ by $a^{\dagger}_{\ep},\
a_{\ep}$.

{}From the above arguments,
at $\mu=0$ we have the enhanced ${\bf Z}_2$ symmetry\foot{
This should be the origin of the ${\bf Z}_2$
symmetry in the tachyon scattering
amplitudes found in \JY (see also \DR \DKR \GTT) as can be seen from the
action $\mu\to -\mu$.}
 and we can indeed
take a quotient by this action. We would like to argue that this
defines the IIA matrix model. After we factor out the
semiclassical part of the wave function by the double-scaling
limit, we get a free relativistic Dirac fermion in the asymptotic
region as in the $c=1$ case \GK \KlR\ before the ${\bf Z}_2$
projection\eqn\fermi{ \Psi_{L}=\sum_{n\in {\bf Z}} a^{L}_n\
e^{in\omega_0(\tau-t)},\ \ \ \Psi_{R}=\sum_{n\in {\bf Z}} a^{R}_n\
e^{-in\omega_0(\tau+t)},} where the `spacial' coordinate
$\tau(\sim \phi)$ is defined by $x \sim \s{2\mu}\cosh(\tau)$. Here
we put a cutoff for a large value of $|\tau|$ and that leads to a
discrete energy $n\omega_0$. The ${\bf Z}_2$ projection identifies
$a_n$ with $a^{\dagger}_{-n}$. After this identification, the
fermion becomes real
\eqn\reall{\Psi_{L,R}^{\dagger}(\tau,t)=\Psi_{L,R}(\tau,t).} After
we take the boundary condition $a^{L}_n=a^{R}_n$ into account, the
only dynamical field in the spacetime is a Majorana fermion.
Indeed, this is the same conclusion as in the previous world-sheet
analysis.

\subsec{Scattering Amplitude from Matrix Model}

The action of the fermion field in the 0A theory
is roughly given by the usual kinetic term
of Dirac fermion plus position dependent Dirac mass term $\sim
\f{1}{\s{M}}e^{-2\tau}
(\Psi_{L}^\dagger\Psi_{L}+\Psi_{R}^\dagger\Psi_{R})$ \NTT \GK.
The point is
that the left and right-moving sector are completely decoupled
with respect to the coordinate $(t,\tau)$. Naively one may think
there is no scattering in such a background of string theory.
However, the space coordinate $\tau$, which is defined by
$x^2=\mu+\s{M+\mu^2}\cosh(2\tau)$, is non-locally related to usual
space coordinate $\phi$ in the string theory as is well-known.
In the asymptotic
region we have $\phi\sim -|\tau|$. Thus a fermion which propagates
from $\tau=-\infty$ to $\tau=\infty$ describes a fermion scattered
off the Liouville potential in string theory. This behavior of
a Majorana fermion agrees with the minisuperspace action \maction\
and the results in section 2.2 at least qualitatively.

In the IIA model, since we have the Majorana projection \reall, we
have the simple fermion action \eqn\action{S=\int dt d\tau (
\Psi_L(\de_t+\de_\tau)\Psi_L+\Psi_R(\de_t-\de_\tau)\Psi_R).} Even
though this is free, the non-local transformation leads to a
reflection at the Liouville potential as we have explained just before.

In the exact wave function analysis of \DKR\ we can get the
non-perturbative S-matrix (or the reflection amplitude) for energy
$\omega$ (in $\al=2$ convention)
\eqn\smat{R_{\omega}=\left(\f{4}{q^2-\f{1}{4}}\right)^{-i\omega}\
\f{\Gamma(\f{1}{2}
-i\omega+\f{|q|}{2})}{\Gamma(\f{1}{2}+i\omega+\f{|q|}{2})}\
e^{i\pi q/2}.} Since this is a pure phase factor, we can conclude
that an incoming RNS fermion is completely reflected at the wall
and the fermion number is conserved. Notice that this scattering
amplitude of fermions takes a rather different form than those of
bosons in bosonic and type 0 string \MPR \five.

To be consistent with the ${\bf Z}_2$ projection \reall\ we should
have $R_{\omega}^*=R_{-\omega}$. This gives an intriguing
quantization $q\in 2{\bf Z}$, which implies that the odd number
$(=q)$ of D0-branes will not be consistent with the ${\bf Z}_2$
orbifold\foot{This may suggest that the original fermion in the 0A
model can be regarded as a sort of a `fractional brane' in IIA.
This fractional object may be related to the spin operator from
the viewpoint of Ising model as the IIA model includes a free
Majorana fermion in the asymptotic region.}. In this case the last
factor $e^{i\pi |q|/2}$ is just $\pm 1$. We can show that the
expansion of \smat\ for the large $|q|$ (or weak coupling) is of
the form
\eqn\expan{R_{\omega}=\pm(1+\sum_{n=1}^{\infty}r_{n}(\omega)q^{-2n}).}
This can be seen from the relation $\f{S(-|q|)}{S(|q|)}
=\f{\sin(\f{1}{2}-\f{i\omega}{2}+\f{|q|}{2})}
{\sin\pi(\f{1}{2}+\f{i\omega}{2}+\f{|q|}{2})}$, where
$S(|q|)\equiv \f{\Gamma(\f{1}{2}
-\f{i\omega}{2}+\f{|q|}{2})}{\Gamma(\f{1}{2}+\f{i\omega}{2}+\f{|q|}{2})}$.
After we neglect the non-perturbative part like $\sim e^{i|q|}$,
we get the expansion \expan. In this way we can see the expected
perturbative expansion of closed string with respect to the string
coupling $g_s^2\sim q^{-2}$. This is non-trivial since in the
bosonic or type0 string we regard a fermion as a D0-brane \MV \KMS
\TT \six\ for which it is in principle possible to have a $g_s
\sim q^{-1}$ expansion.

\newsec{S-matrix of 2D Type IIA String}

The scattering S-matrix
should be one of the most basic quantities when we compare
 a string theory with its dual matrix model. The dynamical field
 in the IIA model is a Majorana fermion as we have observed
 in both the world-sheet theory (section 2)
 and its matrix model dual (section 3). Thus it will be useful to compare
 the scattering amplitudes in both sides. Since we have done in the matrix
 model side in section 3.2, here we would like to analyze those in
 the world-sheet computations. Similar computations have
  been done in \DK \AH\
 for the type 0 string.
 In this section we compute the scattering amplitudes of
 NSR(or RNS) fermions following the method and conventions\foot{Notice that
the paper \DK\ use the $\alpha'=2$ unit and the Liouville field
$\phi$ corresponds to $-\phi$ in our notation defined in section 2.}
 in \DK.
Amplitudes can be written in the following form (notice also that
on-shell NSR(RNS) vertex operators represent incoming(outgoing)
fermions) \eqn\amp{A(p_1,\ddd,p_{M+N}) =q^s \la\ (V_{NSR})^{N}\ \
(V_{RNS})^M\ \ (\int V_{RR})^s\ \lb,} for $N\to M$ scattering with
$s$ insertions of Liouville-like term or RR-flux term given by
\rrl.
 The non-negative
integer $s$
is determined such that the sum of all $\phi$ momenta of the RNS or NSR
vertex operators is given by $-Q-s$.
Even though $s$ becomes a non-integer value for general momenta,
it is natural
to believe
that the general amplitudes are given
by an analytical continuation as usual
in Liouville theory. However,
it is not easy to compute the amplitudes \amp\ for any $s\in {\bf Z}\geq 0$
in a systematical
way due to the presence of the superghost $\vp$ and the
picture changing.

Thus let us first concentrate on the amplitudes which do not
include any insertion of the Liouville-like term \rrl\ (i.e.
$s=0$). Since the three point function is zero due to the
fermionic statics, let us compute the four point function, for
example. To match the picture we consider the correlation function
of the following four vertex operators \eqn\fourp{\la
V^{(-1/2,-1)}_{R(-),NS}(z_1,\bar{z_1})\cdot
V^{(-1/2,0)}_{R(-),NS}(z_2,\bar{z_2})\cdot
V^{(-1,-1/2)}_{NS,R(+)}(z_3,\bar{z_3})\cdot
V^{(0,-1/2)}_{NS,R(+)}(z_4,\bar{z_4}) \lb .} We define the
(Euclidean) momenta of each vertex operators by
$\vec{p}^i=(p_x^i,p_\phi^i)=(k_i,\beta_i)$ for $i=1,2,3,4$. The
on-shell ($L_0=1$) condition is $\beta_i+\f{Q}{2}=|k_i|$ and the
momentum conservation is $\sum_{i}k_i=0$ and $\sum_{i}\beta_i=-Q$.
Also note that the superconformal invariance $G_0=0$ requires
$k_1,k_2<0$ and $k_3,k_4>0$ (called kinematical region). After we
integrate the moduli $z^i$ with the gauge fixing, we get the
following amplitude \eqn\fourv{A(p^1,p^2,p^3,p^4)=-2\pi^4
\left(\f{1}{2}+\vec{p}_2\cdot\vec{p}_4\right)\f{\Gamma(\vec{p}_1
\cdot\vec{p}_4+\f{1}{2})\Gamma(\vec{p}_2
\cdot\vec{p}_4+\f{1}{2})\Gamma(\vec{p}_3 \cdot\vec{p}_4+1)}{
\Gamma(-\vec{p}_1 \cdot\vec{p}_4+\f{1}{2})\Gamma(-\vec{p}_2
\cdot\vec{p}_4+\f{1}{2})\Gamma(-\vec{p}_3 \cdot\vec{p}_4)}.}
Notice that this expression is antisymmetric\foot{To see this note
the identities $\vec{p}_2\cdot\vec{p}_4=
-\vec{p}_1\cdot\vec{p}_4=\vec{p}_1\cdot\vec{p}_3=...$.}
 with respect to
$(\vec{p}_1,\vec{p}_2)$ and $(\vec{p}_3,\vec{p}_4)$ being
consistent with their fermionic statistics. In the kinematical
region $k_1,k_2<0$ and $k_3,k_4>0$, we can show that $\vec{p}_3
\cdot\vec{p}_4=0$. Since we have a divergence from the denominator
and other factors remain finite, we can conclude that that the
amplitude is zero. It is very natural that this result should
extend to higher point functions. Indeed we can show this from a
viewpoint of ground rings\foot{For this we can act the ground ring
elements $x$ and $y$ on the physical vertex operators up to the
BRST cohomology. Especially for the (NS,R) sector vertex, which we
are interested in, we get $x\cdot V_{NS,R}(q)\sim q^2
V_{R,NS}(q+\f{1}{2})$ and $y\cdot V_{NS,R}\sim 0,$ and also an
opposite relation for the RNS vertex. We can move the position of
the operators $x$ or $y$ so that it annihilates a vertex
operator.} applying the arguments \KMSG. Then we can find that the
$s=0$ amplitude is non-zero if and only if we consider $1\to M$ or
 $N\to 1$ scattering. For example, if we return to our previous example
$2 \to 2$ scattering, this is obviously zero in this argument.
Also generally due to the fermionic statistic on the world-sheet,
all of the $1\to M$ and $N\to 1$ scatterings are zero. Thus we
have found that all non-trivial S-matrices are zero if we consider
$s=0$ amplitudes or equally the linear dilaton background ($q=0$).

Now we take the Liouville perturbation \rrl\ into account.
We again consider the four particle scattering
for a positive integer $s$. This was zero when $s=0$ as we
have seen. Let us study the simplest non-trivial example of
$1\to 3$ scattering for $s=1$ . We
assume the momenta of four particles are given by
$k_1>0,k_2<0,k_3<0,k_4<0$. The on-shell condition and momentum
conservation require $k_1=3/2$ and $k_2+k_3+k_4=-3/2$ at the
background charge $Q=2$ for the two dimensional string theory.
Then we get the following integral expression of the amplitude
\eqn\fourp{\eqalign{ &\int dz^2 \la V_{NSR}(\vec{p}_1)(0)\cdot
V_{RNS}(\vec{p}_2)(1)\cdot V_{RNS}(\vec{p}_3)(\infty)\cdot
V_{RNS}(\vec{p}_4)(z)\cdot \left(\int dw^2
V_{RR}(w,\bar{w})\right)\lb \cr
&=-\f{Q(2k_1-Q/2)}{4\s{2}}\left((2k_4+Q/2) I_{1}(k_2,k_4)
+(2k_2+Q/2) I_{2}(k_2,k_4)\right) ,}} where we have defined the
integrals as \eqn\integp{\eqalign{ I_{1}(k_2,k_4)&=\int dz^2 dw^2
|z|^{2\vec{p}_1\cdot \vec{p}_4-1} |1-z|^{2\vec{p}_2\cdot
\vec{p}_4}|w|^{Q\beta_1+1} |1-w|^{Q\beta_2-1}|z-w|^{Q\beta_4-1}
\cr &=\int dz^2 dw^2 |z|^{4k_4} |1-z|^{2k_3+1}|w|^2
|1-w|^{-2k_2-3}|z-w|^{-2k_4-3}, \cr I_{2}(k_2,k_4)&=\int dz^2 dw^2
z^{\vec{p}_1\cdot \vec{p}_4-1/2} \bar{z}^{\vec{p}_1\cdot
\vec{p}_4+1/2} |1-z|^{2\vec{p}_2\cdot \vec{p}_4}w^{Q\beta_1 /2-1}
\bar{w}^{Q\beta_1 /2} |1-w|^{Q\beta_2-1}|z-w|^{Q\beta_4-1} \cr
&=\int dz^2 dw^2 z^{2k_4} \bar{z}^{2k_4+1} |1-z|^{2k_3+1}w
|1-w|^{-2k_2-3}|z-w|^{-2k_4-3}.}} In the computation of the
amplitude \fourp\ we chose the pictures of the five vertex operators as
$(0,-1/2),(-1/2,-1),(-1/2,0),(-1/2,0)$ and $(-1/2,-1/2)$,
respectively.

It is possible to perform these integrals by using the integration
formula found in \FH\ in terms of the generalized hypergeometric
function. As we show the detailed computations in the appendix A,
the scattering amplitude turns out to be vanishing. However, in
this case we cannot easily conclude that the S-matrix is zero for
$s=1$ because we may expect the renormalization of `cosmological
constant' $q$. Indeed in the usual c=1 string, the naive $s>0$
amplitudes are all zero and they become finite after the
renormalization $\mu_{ren}=\lim_{\ep\to 0} \mu \epsilon$. To
examine this correctly we need to regularize the amplitude. We may
start with a general background charge $Q\neq 2$ and a (imaginary)
background charge in the $X$ direction and then take the $Q=2$
limit as we usually do in other two dimensional string models \DKB
\DK. Since the analysis of the integral in this limit is very
difficult, here we want to be satisfied by the evaluation of only
the first integral $I_1$ in \integp. Interestingly, we can show
that $I_1$ is the same amplitude of $1\to 3$ scattering with $s=1$
(in $\al=1$ convention in bosonic string) if we replace $k_i$ with
$k_i+1/2$ for $i=1,2,3$. It is just a constant $c$ times the
leg-factor $\prod_{i}\f{\Gamma(|k_i|)}{\Gamma(-|k_i|)}$ after the
renormalization \DK. Since we know the answer in the two
dimensional bosonic string (by using the matrix model
computation), we get the result \eqn\itt{I_{1}(k_2,k_4)=cq^s
\f{\Gamma(-k_2+1/2)}{\Gamma(k_2-1/2)}\f{\Gamma(-k_3+1/2)}{\Gamma(k_3-1/2)}
\f{\Gamma(-k_4+1/2)}{\Gamma(k_4-1/2)}.} Even though we did not
find the complete total expression for the amplitude, in order to
see its structure this is enough. For example we can easily
speculate from \itt\ the leg-factor in IIA theory
\eqn\legiias{e^{i\delta(\omega)}=q^{-2i\omega}\f{\Gamma(\f{1}{2}+i\omega)}
{\Gamma(\f{1}{2}-i\omega)},} and this agrees with the previous
expectation from the minisuperspace computation \legfm. As the
scattering amplitudes of more than four particles are much more
complicated, we we will not compute explicitly in this paper. The
three particle scattering is obviously zero due to the fermionic
statics.

Finally we discuss the two point function. The trivial two point
function like $\la V_{NSR}V_{NSR}\lb$ and $\la V_{RNS}V_{RNS}\lb$
(trivial $1\to 1$ scattering) is obviously non-zero since this is
the norm of the state\foot{To see this clearly it will be better
to move into Lorentzian frame via $|k|\to -iE$.} in CFT. Also the
non-trivial two point function (or reflection amplitude) is
non-zero as we can be seen from the simplest case $s=1$ of the
amplitude \eqn\refl{\left\la
V_{NSR}\left(k,\beta=k-\f{Q}{2}\right)\cdot
V_{RNS}\left(-k,\beta=k-\f{Q}{2}\right) \left(\int dz^2
V_{RR}(z,\bar{z})\right)^s\right\lb, } which is obviously
non-zero. For general positive integer $s$ the amplitude is
non-vanishing iff $s$ is an odd integer. Thus we find the position
of poles \eqn\ple{k=\f{1}{2},\f{3}{2},\f{5}{2},\ddd.} These
positions of poles agree with the previous results \legfm\ and
\itt.

Before we finish this section, let us summarize the results
obtained in our analysis of string scattering amplitudes. In the
linear dilaton background $q=0$ we have seen that all amplitudes
except the (trivial) two point function are zero. Notice that this
result is different from that of the bosonic or type 0 model,
where we have non-trivial scatterings even when $\mu=0$. In the
presence of a non-zero RR-field $q$, we can find two
possibilities: $(i)$ all amplitudes except the two point
reflection amplitude are zero, or $(ii)$ most of the amplitudes
become non-trivial as in the two dimensional bosonic or type 0
string. On the other hand, we know that in the proposed IIA matrix
model dual, the fermions are free and only the reflection
amplitude ($1\to 1$ scattering) is non-trivial. Thus the
transformation from a fermion $\psi_{mat}$ in the matrix model to
a fermion $\psi_{IIA}$ in IIA string is trivial in the first case
$(i)$. In the second case $(ii)$, however, this will become a
complicated non-local transformation written, for example, in the
following form \eqn\trf{\psi_{mat}\sim e^{i\delta(\omega)}
\psi_{IIA}+a(\omega)\psi_{IIA}\de\psi_{IIA}\de^2\psi_{IIA} +\ddd,}
which changes the fermion number in addition to the presence of
the leg factor. This is similar to the bosonization of fermions
that appear in the bosonic or type 0 string case.

Even though we found that the $1\to 3$ scattering amplitude with
$s=1$ is zero, a more careful analysis assuming a possible
renormalization of $q$ suggests the second possibility $(ii)$.
This was because the structures of this amplitude looks similar to
that of two dimensional bosonic or type 0 string. This possibility
is also more natural since we usually expect back-reactions from
the gravity sector which will lead to the non-zero four point
functions\foot{We would like to thank A. Strominger for pointing
out this point.}. However, the above arguments are not conclusive
and both possibilities may be possible within our results. In
order to go beyond this we need a more systematic method to
compute the scattering amplitudes in the presence of RR-flux.

\newsec{Conclusions}

In this paper we discussed the type IIA string theory in two
dimension. We presented both the world-sheet description and its
matrix model dual. This model describes a two dimensional
spacetime with a Majorana fermion coupled to gravity. The matrix
model description shows that it is non-perturbatively stable. We
argued that this model can be weakly coupled due to the background
RR-flux instead of the familiar cosmological constant. We also
examined tree level scattering amplitudes on the world-sheet
theory and compared them with the matrix model. Even though we did
not determine them completely in the world-sheet computations, we
found intriguing structures in the IIA amplitudes, which cannot be
found in the bosonic or type 0 string. We will leave further
analysis of the scattering amplitudes for a future problem. In
order to compute such physical quantities in a systematical way, a
construction of Green-Schwarz-like formalism may be useful. It
will also be an interesting question how D-branes are related to
this matrix model.

\centerline{\bf Acknowledgments}

I would like to thank A. Strominger for extremely helpful comments
and suggestions. I am also grateful to T. Eguchi, D. Gaiotto, D.
Ghoshal, S. Gukov, J. Karczmarek,  E. Martinec, J. McGreevy, S.
Minwalla, S. Mukhi, Y. Nakayama, Y. Sugawara, H. Takayanagi, S.
Terashima, N. Toumbas and S. Yamaguchi
 for useful discussions. The work was supported in part by DOE
grant DE-FG02-91ER40654.

\appendix{A}{The proof of the vanishing of the four point amplitude}

Here we would like to show that the direct evaluations of
integrals $I_1$ and $I_2$ in \fourp\ are zero. First let us prove
explicitly that \eqn\zesh{ I_{2}(k_2,k_4)=\int dz^2 dw^2 z^{2k_4}
\bar{z}^{2k_4+1} |1-z|^{2k_3+1}w |1-w|^{-2k_2-3}|z-w|^{-2k_4-3},}
is vanishing. The similar direct computation can be applied to
show $I_1=0$ as one can easily see.

The more general form of the integral \eqn\ccint{I=\int dz^2dw^2
z^{\ap_1} (1-z)^{\ap_2}
\bar{z}^{\bar{\ap_1}}(1-\bar{z})^{\bar{\ap_2}} w^{\ap'_1}
(1-w)^{\ap'_2} \bar{w}^{\bar{\ap'_1}}
(1-\bar{w})^{\bar{\ap'_2}}|z-w|^{4\sigma},} was computed in \FH\
(see appendix of that paper) in terms of the generalized
hypergeometric function
$_3F_2$. The result of the integral is given by (we use the expression
found in \SA ) \eqn\reint{I=D_1C^{12}(\ap)C^{12}(\bar{\ap})+ D_2
C^{21}(\ap)C^{21}(\bar{\ap})+D_3\{C^{12}(\ap)C^{21}(\bar{\ap})
+C^{21}(\ap)C^{12}(\bar{\ap})\},} where $C^{ab}(\ap)\ \
(a,b=1,2,3)$ is defined by ($C^{ab}(\bar{\ap})$ can be obtained by
replacing $\ap_a$ and $\ap'_b$ with $\bar{\ap}_a$ and
$\bar{\ap'}_b$) \eqn\defff{\eqalign{
C^{ab}(\ap)=&\f{\Gamma(1+\ap_a+\ap'_a-k')\Gamma(1+\ap_b+\ap'_b-k')
\Gamma(1+\ap'_a)\Gamma(1+\ap_b) } {\Gamma(\ap'_a-\ap_c+1)
\Gamma(\ap_b-\ap'_c+1)} \cr &\cdot
{}_3F_2(1+\ap'_a,1+\ap_b,k'-\ap_c-\ap_c';
\ap'_a-\ap_c+1,\ap_b-\ap'_c+1;1).}} We also defined $\ap_3$ and
$k'$ by $\ap_1+\ap_2+\ap_3+1=k'=-2\sigma-1$. The functions $D_i$\
are given by \eqn\funcd{\eqalign{ D_1=&\f{s(\ap'_1)s(\ap'_2)
\left(
s(\ap_1)s(\ap'_1)s(\ap_3)-s(\ap'_3)s(\ap_1-k')s(\ap_3-\ap'_2)\right)
} {s(\ap_3)s(\ap'_3)s(\ap_3+\ap_3'-k')}, \cr
D_3=&-\f{s(\ap_1)s(\ap'_1)s(\ap_2)s(\ap'_2)s(\ap_3+\ap'_3)}
{s(\ap_3)s(\ap'_3)s(\ap_3+\ap_3'-k')},}} where $s(\ap)\equiv
\sin(\pi\ap)$. $D_2$ can be obtained by replacing $\ap_a$ with
$\ap'_a$. The integral \zesh\ corresponds to the values
$\ap_1=\bar{\ap_1}-1=2k_4,\ \ap_2=\bar{\ap_2}=k_3+1/2,\
\ap_1'=\bar{\ap_1'}-1=0,\ \ap'_2=\bar{\ap'_2}=-k_2-3/2,\
k'=k_4+1/2$. Since $\ap'_1$ is an integer, we can find
$D_1=D_3=0$. Thus we have only to compute $C^{21}(\ap)$ and
$C^{21}(\bar{\ap})$. $C^{21}(\ap)$ can be obtained as follows
\eqn\estcom{\eqalign{C^{21}(\ap)=&\f{\Gamma(2k_4+1)\Gamma(2k_3+1)
\Gamma(-k_2-1/2)
\Gamma(-k_4-1/2)}{\Gamma(-2k_2-1)\Gamma(k_3+1/2)} \cr &
\cdot{}_3F_2(-k_2-1/2,2k_3+1,-1;-2k_2-1,k_3+1/2;1).}} As the one
of the parameters of ${}_3F_2$ is a negative integer $-1$, the
hypergeometric function is just a sum of two terms. Easily we can
see
\eqn\zee{{}_3F_2(-k_2-\f{1}{2},2k_3+1,-1;-2k_2-1,k_3+\f{1}{2};1)
=1-\f{(k_2-\f{1}{2})(2k_3-1)}{(2k_2-1)(k_3-\f{1}{2})}=0.} Thus we
have shown $C^{21}(\ap)=0$. In the same way we can compute
$C^{21}(\bar{\ap})$ and we find a finite value \eqn\ffgf{
C^{21}(\bar{\ap})=\f{\Gamma(2k_4+2)\Gamma(2k_3+1)\Gamma(-k_4-1/2)
\Gamma(-k_2-1/2)}{\Gamma(-2k_2)\Gamma(k_3+1/2)}.} In conclusion we
have found $I_2(k_2,k_4)=D_2C^{21}(\ap)C^{21}(\bar{\ap})=0$.

Another integral $I_1$ in \fourp\ can also be computed in a similar
way. Indeed we can see that $I_1(k_2,k_4)=D_2(C^{21}(\ap))^2=0$,
where the values of $\ap_a$ are the same as before.

\listrefs

\end